\documentclass{iopart}
\usepackage{graphicx}
\usepackage{latexsym}
\usepackage{amssymb}
\usepackage{verbatim}

\begin{document}
\newcommand{\musr}{$\mu$SR{}}
\newcommand{\chem}[1]{\ensuremath{\mathrm{#1}}}

\title{Muon spin relaxation study of LaTiO$_3$ and YTiO$_3$}

\author{P.\ J.\ Baker$^1$, T.\ Lancaster$^1$, S.\ J.\ Blundell$^1$, W.\ Hayes$^1$, 
\\ F.\ L.\ Pratt$^2$, M.~Itoh$^3$, S.~Kuroiwa$^4$ and J.~Akimitsu$^4$}
\address{$^1$ Department of Physics, Oxford University, 
Parks Road, Oxford OX1 3PU, United Kingdom}



\address{$^2$ ISIS Muon Facility, Rutherford Appleton Laboratory, 
Harwell Science and Innovation Campus, Didcot, OX11 0QX, United Kingdom}


\address{$^3$ Department of Physics, Nagoya University, Nagoya 464-8602, Japan}

\address{$^4$ Department of Physics, Aoyama Gakuin University, Sagamihara, Kanagawa 229-8558, Japan}

\ead{p.baker1@physics.ox.ac.uk}

\date{\today}

\begin{abstract}
We report muon spin relaxation ($\mu$SR) measurements on two \chem{Ti^{3+}} containing perovskites, \chem{LaTiO_{3}} and \chem{YTiO_{3}}, which display long range magnetic order at low temperature. 
For both materials, oscillations in the time-dependence of the muon polarization are observed 
which are consistent with three-dimensional magnetic order. From our data we identify two 
magnetically inequivalent muon stopping sites. The $\mu$SR results 
are compared with the magnetic structures of these compounds previously derived from neutron 
diffraction and $\mu$SR studies on structurally similar compounds. 
\end{abstract}

\submitto{JPCM}
\maketitle

\section{\label{sec:Introduction} Introduction}
\begin{figure}[htb]
\begin{center}
\includegraphics[width=13cm]{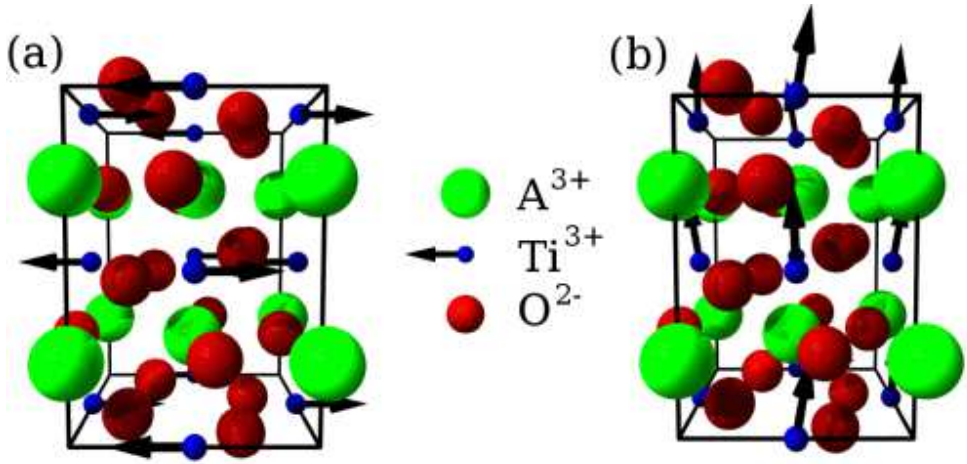}
\end{center}
\caption{(a) \chem{LaTiO_3} and (b) \chem{YTiO_3}, showing the 
magnetic structures previously proposed. Structural parameters were taken from 
Refs.~\cite{PhysRevB.68.060401}~\&~\cite{maclean79}, and magnetic structures from 
Refs.~\cite{PhysRevB.68.060401}~\&~\cite{PhysRevLett.89.167202}.
} 
\label{fig:xto:structure}
\end{figure}

Despite their structural simplicity (exemplified in Figure~{\ref{fig:xto:structure}}) 
perovskite compounds of the form \chem{ABX_3} show a wide variety of physical 
properties, particularly when the simple cubic structure is distorted~\cite{cox}. 
Changing the ionic radius of the ion on the \chem{A} site allows the 
distortion to be controlled and, through this, the physics of these 
materials can be tuned~\cite{PhysRevB.75.224402}. 
An example of two similar compounds where a small change in the ionic radius 
causes a significant change in the physical properties is the pair 
\chem{LaTiO_3} and \chem{YTiO_3}. 

These two compounds are Mott-Hubbard insulators but retain the orbital degree of freedom 
in the $t_{2g}$ state~\cite{PhysRevB.74.054412} and show a strong coupling between
spin and orbital degrees of freedom~\cite{NJP.6.154}.  Orbital
degeneracy, which can lead to phenomena such as colossal
magnetoresistance or unconventional
superconductivity~\cite{Science.288.462}, is present in isolated
\chem{Ti} $t_{2g}$ ions, but is lifted in these
compounds~\cite{NJP.6.154}.  The size of the \chem{A^{3+}} ion provides
one means of tuning the properties of these 
titanates~\cite{NJP.6.154}, affecting the \chem{Ti}-\chem{O}-\chem{Ti}
bond angles and exchange interactions.  This is evident in the difference between 
the low temperature magnetic structures of these two compounds, observed using neutron 
diffraction~\cite{PhysRevLett.85.3946,PhysRevB.68.060401,PhysRevLett.89.167202}. 
\chem{LaTiO_3} is a G-type antiferromagnet with the \chem{Ti} moments aligned along the 
$a$-axis~\cite{PhysRevLett.85.3946,PhysRevB.68.060401} below $T_{\rm N}$. The precise 
value of $T_{\rm N}$ is very sensitive to the oxygen stoichiometry and reports vary between 
$120$ and $\sim 150$~K~\cite{PhysRevLett.97.157401}. \chem{YTiO_{3}} orders
ferromagnetically~\cite{PhysRevLett.89.167202} with the spins aligned
along the $c$-axis at $T_{\rm C} = 27$~K; however, there is a G-type
antiferromagnetic component along $a$, and an A-type component along
$b$ (see figure~\ref{fig:xto:structure}).

Evidence of orbital excitations due to fluctuations of orbital-exchange bonds 
has been found in \chem{LaTiO_3} and \chem{YTiO_3} using Raman scattering, and these 
excitations are remarkably similar to the exchange-bond fluctuations which give rise to 
magnetic Raman scattering in cuprates~\cite{PhysRevLett.97.157401}. A broad range of 
measurements have demonstrated the underlying orbital ordering in both compounds~\cite{PhysRevB.75.224402,PhysRevB.68.060401,PhysRevLett.94.056401,PhysRevLett.91.066403,JPSJ.70.3475,PhysRevLett.93.257207}, strongly excluding the orbital liquid picture hypothesized for \chem{LaTiO_3}~\cite{PhysRevLett.85.3950} and agreeing with the reduced orbital moment 
found in X-ray and NMR measurements on \chem{LaTiO_3}~\cite{PhysRevLett.85.3946,PhysRevLett.91.167202}.
It has been shown~\cite{PhysRevB.71.184431} that the \chem{Y_{1-{\it x}}La_{\it x}TiO_3} 
system is an itinerant-electron antiferromagnet with no orbital ordering for $x >0.7$ and that an intermediate phase exists for $0.3 < x < 0.7$, with orbital-order fluctuations and ferromagnetic interactions that reduce $T_{\rm N}$. For $x < 0.3$ the system shows orbital ordering and a ferromagnetic transition and it was suggested that even at $x=0$ the volume of the orbitally ordered region does not encompass the whole sample. 

Theoretical work on these compounds has focused around the mechanism
that selects the ground state from the possible spin and orbital
configurations. Models considering the orbitals as quasi-static
entities
\cite{PhysRevB.74.054412,PhysRevB.68.060401,PhysRevLett.91.167203,PhysRevLett.92.176403,PhysRevB.71.144412}
satisfactorily predict the orbital occupation and magnetic ordering.
Nevertheless, there remain aspects of the experimental
observations~\cite{PhysRevLett.85.3946,PhysRevLett.89.167202,PhysRevLett.97.157401}
that cannot be successfully described without including the quantum
fluctuations of the
orbitals~\cite{PhysRevLett.85.3950,PhysRevLett.89.167201,PhysRevB.68.205109}, 
particularly with regard to the Raman scattering results. 
With quasi-static orbital occupations, excitations are in the form of well-defined 
crystal field excitations, whereas if fluctuations are significant, the 
excitations are collective modes, and it is the latter which are observed 
by Raman scattering experiments~\cite{PhysRevLett.97.157401}. 
Predicting the magnetic properties of these compounds based on their structures (i.e. the tuning 
provided by the $A$-site cation radius) and their observed orbital physics has proved challenging, 
particularly for \chem{LaTiO_3}~\cite{PhysRevB.74.054412}. In this context, additional 
detailed characterisation of the magnetic properties of both compounds is worthwhile, in the 
hope of providing information to further constrain the theoretical models.

In this paper we describe the results of a muon-spin relaxation ($\mu$SR) investigation 
into the magnetic properties of \chem{LaTiO_3} and \chem{YTiO_3}. The methods of 
synthesis and the experimental details common to both compounds are explained in 
section~{\ref{sec:xto:exp}}. The results of the $\mu$SR experiments 
are presented in sections~{\ref{sec:xto:lto}}~and~{\ref{sec:xto:yto}}. Dipole 
field calculations for magnetic structures previously deduced by neutron diffraction 
are compared to the $\mu$SR results in section~{\ref{sec:xto:dipole}}. The results 
are discussed and conclusions are drawn in section~{\ref{sec:xto:discussion}}.

\section{\label{sec:xto:exp} Experimental}

The \chem{LaTiO_3} sample was synthesized by arc melting appropriate
mixtures of \chem{La_{2}O_{3}}, \chem{TiO_2}, and \chem{Ti} in an
argon atmosphere~\cite{JPSJ.68.2783}. The properties of \chem{LaTiO_3}
are strongly dependent on the oxygen stoichiometry (see, for examples,
Refs.~\cite{PhysRevLett.97.157401,PhysRevB.71.184431}). To produce a sample as 
close to the correct stoichiometry as possible, several samples were prepared and 
one with $T_{\rm N} = 135$~K, determined by magnetic measurements, was chosen. 
The \chem{YTiO_3} was prepared similarly, using \chem{Y_{2}O_{3}}, and was 
determined to be \chem{YTiO_{3+\delta}} with $\delta \leq 0.05$, 
$T_{\rm C} = 27$~K, and a saturation magnetic moment of 
$0.84 \mu_{\rm B}/$\chem{Ti}~\cite{JPSJ.70.3475}. 

\begin{figure}[htb]
\begin{center}
\includegraphics[width=\linewidth]{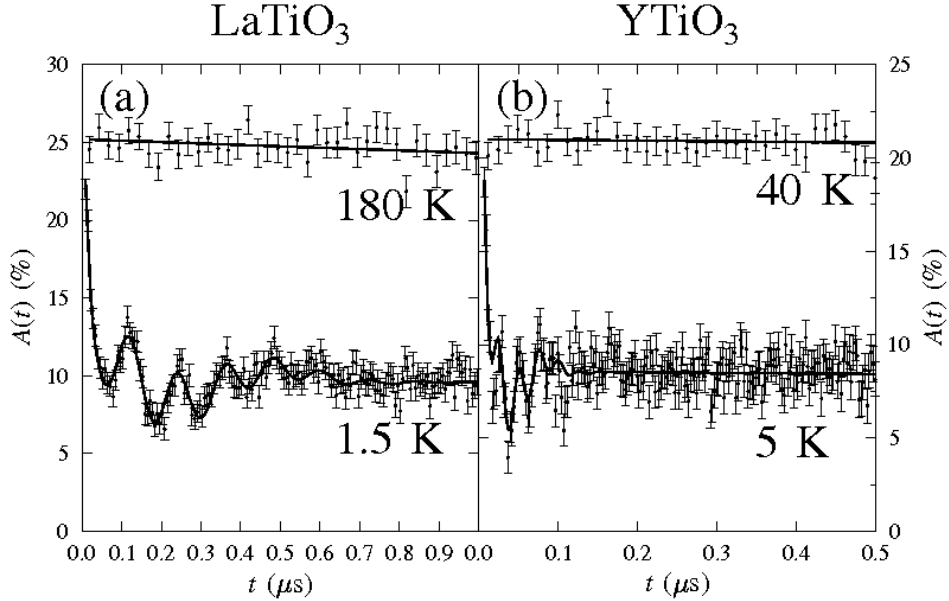}
\end{center}
\caption[Asymmetry spectra for LaTiO$_3$]{
Examples of the raw \musr{} data recorded for (a) \chem{LaTiO_3} 
and (b) \chem{YTiO_3}. 
For both compounds, the precession is clearly evident in the low temperature data and 
absent in the high temperature data.
For the low-temperature datasets the lines plotted are fits of the data to Equation~{\ref{eq:xto:xtofit}}, and for the high-temperature datasets the lines are 
fits to an exponential relaxation, as discussed in the text.
}
\label{fig:xto:xtodata}
\end{figure}

Our $\mu$SR experiments on both samples were carried out using the 
GPS instrument at the Paul Scherrer Institute, in zero
applied magnetic field (ZF). In a $\mu$SR experiment~\cite{blundell99} 
spin polarized positive muons are implanted into the sample, 
generally stopping at an interstitial position within the crystal structure, 
without significant loss of polarization. The polarization, 
$P_{z}(t)$, of the muon subsequently depends on the magnetic environment 
of the stopping site and can be measured using the asymmetric decay of 
the muon, with around 20 million muon decays recorded for each temperature 
point considered. The emitted positron is detected in scintillation 
counters around the sample position~\cite{blundell99}. The asymmetry of the 
positron counts is $A(t)=(A(0)-A_{\rm bg})P_{z}(t) + A_{\rm bg}$, with 
$A(0) \sim 25$~\% (see Figure~{\ref{fig:xto:xtodata}}) and $A_{\rm bg}$ a 
small contribution to the signal due to muons stopping outside the sample. 
The polycrystalline samples were wrapped in silver foil packets and mounted on 
a silver backing plate, since the small nuclear magnetic moment of silver minimizes 
the relaxing contribution of the sample mount to $A_{\rm bg}$. Examples of the measured 
asymmetry spectra in both compounds are presented in Figure~{\ref{fig:xto:xtodata}}.  
At low temperature, precession signals are seen in both compounds, indicative
of long-range magnetic order, with two precession frequencies (see 
Figures~\ref{fig:xto:ltoresults}~and~\ref{fig:xto:ytoresults})
indicating two magnetically inequivalent muon sites. Above their
respective transition temperatures the data for both compounds shows
exponential relaxation characteristic of a paramagnetic phase. 

After the initial positron decay asymmetry, $A(0)$, and the background, $A_{\rm bg}$, 
had been determined, the following equation was used to analyse the asymmetry 
data below the magnetic ordering temperature in each compound:
\begin{equation}
P_{z}(t) = P_{\rm f} e^(-\lambda t) + P_{\rm r} e^{-\sigma^{2}_{\rm r} t^2} + P_{\rm osc} e^{-\sigma^{2}_{\rm osc} t^2}[\cos(2\pi \nu_1 t) + \cos(2\pi \nu_2 t)].
\label{eq:xto:xtofit}
\end{equation}
The components $P_{\rm f}$, $P_{\rm r}$, and $P_{\rm osc}$ are all independent of temperature 
and are in the ratio $(P_{\rm f} + P_{\rm r}) / P_{\rm osc} \simeq 2$ expected from 
polycrystalline averaging. The exponentially relaxing component $P_{\rm f}$ can be attributed 
to fluctuating fields parallel to the direction of the implanted muon spin, and the relaxation 
rate, $\lambda$ was found to be almost independent of temperature. A Gaussian relaxing component, 
$P_{\rm r}$, describes the rapid drop in the asymmetry at short times, due to large magnetic 
fields at a muon stopping site, and the $P_{\rm osc}$ term describes the two-frequency 
oscillating component of the signal due to coherent local magnetic fields at two magnetically 
inequivalent muon stopping sites (we take $\nu_1 > \nu_2$). The data were fitted 
throughout the ordered temperature range while fixing the ratio $\nu_{2}/\nu_{1}$ 
to the value obtained at base temperature. For both compounds the function
\begin{equation}
\nu_{i}(T)=\nu_{i}(0)(1-(T/T_{\mathrm c})^{\alpha})^{\beta}
\label{eq:xto:nuoft}
\end{equation}
was used to fit the temperature dependences of the precession frequencies $\nu_{i}(T)$, 
where $T_{\mathrm c}$ is the appropriate ordering temperature, $\alpha$ describes the 
temperature dependence as $T \rightarrow 0$, and $\beta$ is the critical parameter 
describing the sublattice magnetization close to $T_{\mathrm c}$~\cite{blundell01}. 

\section{\musr{} measurements on LaTiO$_3$}
\label{sec:xto:lto}

\begin{figure}[htb]
\begin{center}
\includegraphics[width=13cm]{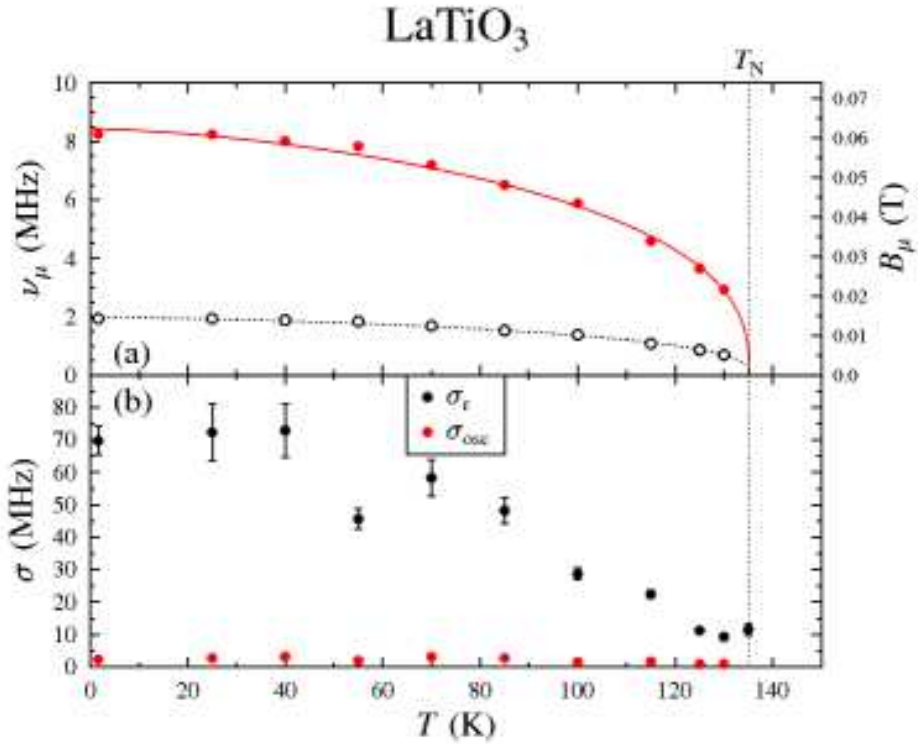}
\end{center}
\caption[Parameters extracted from LaTiO$_3$ \musr{} data]{
Parameters extracted from the raw \musr{} data on \chem{LaTiO_3} using 
Equation~{\ref{eq:xto:xtofit}}: 
(a) Precession frequencies $\nu_1$ and $\nu_2$, together with the 
equivalent magnetic field. 
(b) Gaussian relaxation rate and linewidth, $\sigma_{\rm r}$ and $\sigma_{\rm osc}$.
Fitted lines in (a) are to Equation~\ref{eq:xto:nuoft} with the parameters discussed in 
the text.
}
\label{fig:xto:ltoresults}
\end{figure}

Raw data recorded on \chem{LaTiO_3} are shown in Figure~{\ref{fig:xto:xtodata}}(a).
The high temperature data are well described by a single exponential relaxation 
consistent with fast fluctuating electronic moments in the paramagnetic phase. 
Muon precession is clearly evident in the ordered phase. 
The fits shown in Figure~{\ref{fig:xto:xtodata}}(a) were to equation~{\ref{eq:xto:xtofit}}.
The ratio $\nu_2 / \nu_1$ was set to $0.234$ from the base temperature data. We see that the 
precession is rapidly damped in the ordered phase since the linewidth is comparable 
to the precession frequencies. The parameters obtained from fitting Equation~{\ref{eq:xto:xtofit}} 
to the asymmetry data, applying these constraints, are shown in Figure~{\ref{fig:xto:ltoresults}}.

Both precession frequencies shown in Figure~{\ref{fig:xto:ltoresults}}(a) 
are well defined up to $T_{\rm N}$, although it was not possible to resolve a 
precession signal in the $135$~K dataset even though the fast relaxing component 
was still evident at this temperature, whereas $A(t)$ for $T \geq 140$~K took the 
simple exponential form expected for a fast-fluctuating paramagnetic phase. 
The values of $\nu_1$ were fitted to Equation~\ref{eq:xto:nuoft} with $\alpha = 1.5$, 
leading to the parameters $\nu_1(0) = 8.4(1)$~MHz, $\beta = 0.37(3)$, 
and $T_{\rm N} = 135(1)$~K. This value of $T_{\rm N}$ is consistent with the value 
found by Zhou and Goodenough~\cite{PhysRevB.71.184431}, and it is conceivable that 
other magnetic studies may have been strongly affected by small regions with 
slightly different oxygen stoichiometry, giving the appearance of a slightly 
higher $T_{\rm N}$. The linewidth of the oscillating components, $\sigma_{\rm osc}$, 
is close to being temperature independent, $\sim 2$~MHz. The Gaussian relaxation rate 
$\sigma_{\rm r}$ is significantly larger than either of the precession frequencies, 
and roughly scales with the precession frequencies, suggesting that muons are stopping 
at sites with very large local fields, probably sitting along the magnetic moment 
direction of nearby \chem{Ti^{3+}} ions. 

\section{\musr{} measurements on YTiO$_3$}
\label{sec:xto:yto}

\begin{figure}[htb]
\begin{center}
\includegraphics[width=13cm]{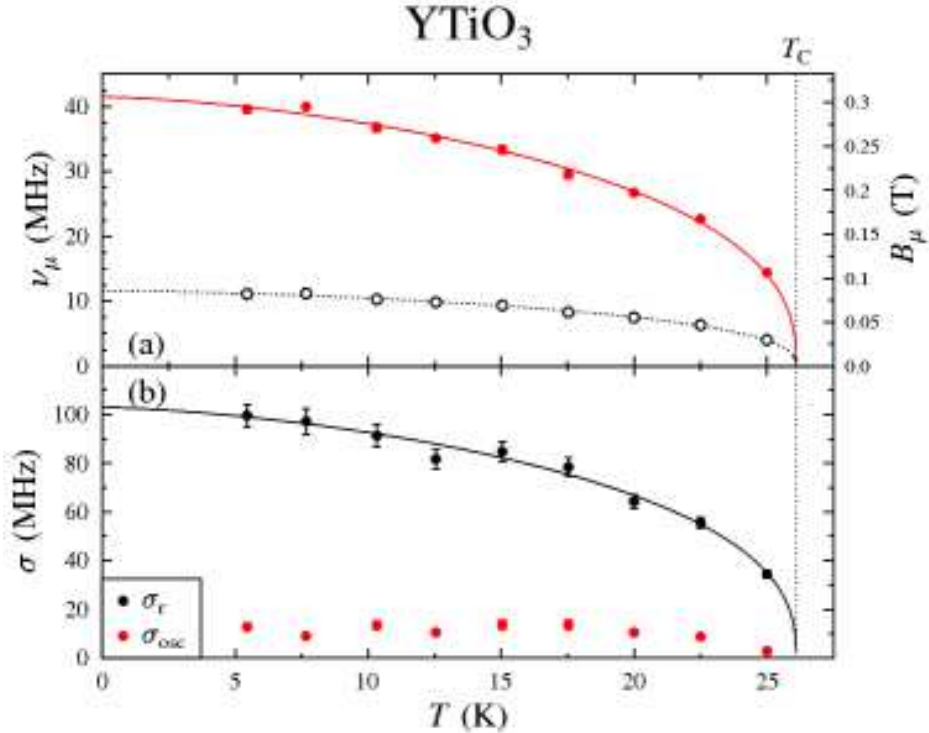}
\end{center}
\caption[Parameters extracted from YTiO$_3$ \musr{} data]{
Parameters extracted from the raw \musr{} data on \chem{YTiO_3} using 
Equation~{\ref{eq:xto:xtofit}} as discussed in the text. 
(a) Precession frequencies $\nu_1$ and $\nu_2$, together with the equivalent 
magnetic field. 
(b) Gaussian relaxation rate and linewidth, $\sigma_{\rm r}$ and $\sigma_{\rm osc}$.
Fitted lines are to Equation~\ref{eq:xto:nuoft} with the parameters discussed in 
the text.
}
\label{fig:xto:ytoresults}
\end{figure}

Asymmetry spectra recorded on \chem{YTiO_3} are shown in Figure~{\ref{fig:xto:xtodata}}(b). 
Again, the high temperature data are well described by a single exponentially relaxing 
component, as is typical for paramagnets. Below $T_{\rm C} \sim 27$~K two muon precession 
frequencies are again observed, consistent with long range magnetic order developing below 
this temperature. 
Preliminary fitting showed that the amplitude of each component of Equation~\ref{eq:xto:xtofit} 
was essentially temperature independent below $T_{\rm C}$, with 
$(P_{\rm f} + P_{\rm r}) / P_{\rm osc} \simeq 2$, and well defined. 
The ratio $\nu_2 / \nu_1$ was set to the ratio at base temperature, $0.28$.
The fits to the data shown in Figure~{\ref{fig:xto:xtodata}}(b) are to 
Equation~{\ref{eq:xto:xtofit}} with the parameters shown in Figure~{\ref{fig:xto:ytoresults}}.

The two precession frequencies shown in Figure~{\ref{fig:xto:ytoresults}}(a) 
remain in proportion for all temperatures below $T_{\rm C} = 27$~K. Unlike the 
situation in \chem{LaTiO_3} however, we see that the fast-relaxing Gaussian 
component has a rate $\sigma_{\rm r}$ which follows a similar power law to the 
precession frequencies. In \chem{YTiO_3} the values of $\nu_1$ and $\sigma_{\rm r}$ 
determined independently in the analysis of the asymmetry data were found to be 
proportional to one another, in agreement with the model of muon sites with very 
large local fields suggested above, so both were fitted to Equation~\ref{eq:xto:nuoft} 
in parallel, fixing $\alpha = 1.5$, leading to the parameters $\nu_1(0) = 41(1)$~MHz, 
$\sigma_{\rm r}(0) = 103(2)$~MHz, $\beta = 0.39(4)$, and $T_{\rm C} = 26.0(4)$~K. 
The linewidth of the oscillating components is $\sim 10$~MHz at low temperature, 
falling slightly towards the transition.

\section{Dipole field calculations}
\label{sec:xto:dipole}

The magnetic structures of \chem{LaTiO_3} and \chem{YTiO_3} have previously been determined 
using neutron scattering~\cite{PhysRevLett.85.3946,PhysRevB.68.060401,PhysRevLett.89.167202}, 
although there remained some uncertainty over the orientation of the magnetic moments 
in \chem{LaTiO_3}~\cite{PhysRevB.68.060401}. These magnetic structures can be 
compared with the $\mu$SR data by calculating the dipolar fields:
\begin{equation}
B_{\rm dip}({\bf r}_{\mu}) = \frac{\mu_0}{4\pi} \sum_i \frac{3({\mathbf \mu}_{i} \cdot 
{\bf \hat{n}}_i){\bf \hat{n}}_i-{\mathbf \mu}_{i}}{\vert {\bf r}_{\mu} - {\bf r}_i \vert ^3},
\label{eq:exp:dipolefield}
\end{equation}
where ${\bf r}_{\mu}$ is the position of the muon, ${\mathbf \mu}_{i}$ is the ordered magnetic 
moment of the $i$th \chem{Ti} ion and 
${\bf \hat{n}}_i (=({\bf r}_{\mu} −{\bf r}_i )/\vert {\bf r}_{\mu} - {\bf r}_i \vert)$ is 
the unit vector from the \chem{Ti} ion at site ${\bf r}_i$ to the muon for points within the unit cell. 
Contributions from of order $10^4$ unit cells were considered. Of course, this method neglects the 
hyperfine contact field, the Lorentz field and the demagnetizing field, although the latter two are zero for antiferromagnets and the contribution of the former to the magnetic field experienced at muon stopping sites, $\sim 1$~\AA{} from \chem{O^{2-}} ions, is generally small. The details specific to each compound will be discussed in sections~{\ref{subsec:dipolelto}}~and~{\ref{subsec:dipoleyto}} below.

Such dipole field calculations have been compared to \musr{} data in other perovskite 
compounds. Some of the more thoroughly studied materials have been the rare 
earth orthoferrites, \chem{{\it R}FeO_3}. The \chem{{\it R} = Sm, Eu, Dy, Ho, Y}, 
and \chem{Er} variants were studied by Holzschuh {\it et al.}~\cite{holzschuh83} 
and they found that the stable muon site common to all of these compounds was on 
the mirror plane at $z = 1/4\,(3/4)$, this being the rare earth--oxygen layer, 
either about $1$ or $1.6$~\AA~from the nearest oxygen ion, as would be expected for 
the \chem{(OH)^-} analog, \chem{(O\mu)^-}. This study was followed by others taking 
a slightly different approach to finding the muon sites~\cite{boekema84,lin86}, 
and these found further plausible sites, albeit apparently metastable ones, 
neighbouring the rare earth--oxygen layers.
Results of these studies have also been applied to orthorhombic nickelates, without 
precession frequencies to test the hypothesis, but the approach was consistent with 
phase separation occurring within magnetically inequivalent layers~\cite{garcia95}.
The most immediately relevant example within the literature is \chem{LaMnO_3}~\cite{cg01}, 
for which a detailed study showed that the two observed precession frequencies corresponded 
to two structurally inequivalent muon sites, the lower frequency one within the rare 
earth--oxygen mirror plane and the higher frequency one at an interstitial site 
within the \chem{Mn}--\chem{O} plane. The latter site requires a significant 
contribution from the contact fields due to the neighbouring oxygen ions, which 
the dipole field calculations presented here do not consider.

\subsection{\chem{LaTiO_3}}
\label{subsec:dipolelto}

Dipole field calculations were carried out for the $G$-type magnetic structure 
reported in Ref.~\cite{PhysRevLett.85.3946} and shown in Figure~{\ref{fig:xto:structure}}(a), 
assuming the magnetic moments ($\mu = 0.57\,\mu_{\rm B}$) are aligned along the 
$a$-axis~\cite{PhysRevB.68.060401}. Calculations were also carried out assuming 
alignment along the $c$-axis as this possibility had previously been favoured and neutron 
measurements did not clearly exclude it~\cite{PhysRevLett.85.3946,PhysRevB.68.060401}. 
The results are periodic in the $c$-axis by half the orthorhombic $c$-axis lattice constant. 
We would expect the muon sites to lie within the $z = 1/4$ plane, as they do in 
\chem{LaMnO_3}~\cite{cg01}. If the moments are along the $c$-axis, the only 
contours corresponding to both observed precession frequencies are very closely 
spaced at points around $0.75$~\AA~from the oxygen ion centres within the plane. 
For moments aligned along the $a$-axis the calculations give results much more 
similar to those in \chem{LaMnO_3}. 
Since we expect the \chem{O}--$\mu$ bond to be around $1$~\AA, this moment orientation 
seems far more consistent with the observed precession frequencies. The other 
possibility is that the muon sites lie within the \chem{Ti}--\chem{O} layer. This 
is far more consistent with moment alignment along the $c$-axis, since suitable 
field values are found at sites between oxygen ions. It is more difficult to make 
precise assignment of muon sites in this case because the field contours are far 
more closely spaced. While there remains some ambiguity, observing well separated 
field contours corresponding to previously identified muon sites for similar materials, 
and apparently equally numbers of plausible muon sites for each frequency, in agreement 
with the experimental amplitudes, is strong evidence that the moments are aligned 
along $a$ rather than $c$, something neutron results have not been able to 
demonstrate with more certainty~\cite{PhysRevB.68.060401}. 

\subsection{\chem{YTiO_3}}
\label{subsec:dipoleyto}

Dipole field calculations were carried out for the ferromagnetic structure 
reported in Ref.~\cite{PhysRevLett.89.167202}, with moment values of 
($0.106,0.0608,0.7034)\,\mu_{\rm B}$ along the principal axes of the pseudocubic 
unit cell ($a,b,c$), and depicted in Figure~{\ref{fig:xto:structure}}(b). 
The calculations show that the magnetic fields for this largely ferromagnetic 
structure are much greater than those in the antiferromagnetic structure of \chem{LaTiO_3}. 
As in \chem{LaTiO_3} the lower frequency component in the signal is consistent with sites 
within the \chem{A}--\chem{O} plane ($z=1/4$), but there are no sites within this layer 
that would correspond to the higher frequency observed. The higher frequency component 
appears consistent with a smaller number of sites between oxygen ions near to or in the 
$z=1/2$ layer, but rather closer to the \chem{Ti^{3+}} ion positions. There are also 
plausible sites corresponding to the lower frequency within this layer. Because of the 
small magnetic moments along the $a$ and $b$-axes the contours are more distorted than 
those calculated for \chem{LaTiO_3}. Considering the variation of these distortions along 
the $c$-axis leads to a structure not dissimilar to a helically ordered magnet, for these 
small components. This could also lead to structurally equivalent sites with much higher 
local fields but significant magnetic inequivalencies, leading to the fast-relaxing 
component, $P_{\rm r}$, of the observed asymmetry. 

\section{Discussion}
\label{sec:xto:discussion}

The \musr{} results clearly demonstrate intrinsic magnetic order below the expected ordering 
temperatures in both samples. We are also able to follow the temperature dependence of the 
(sub)lattice magnetization and show that the behaviour is essentially conventional. The 
values of $\beta$ derived from Equation~\ref{eq:xto:nuoft} describe the behaviour close 
to the transition temperature. The values of $\beta = 0.37$ (\chem{LaTiO_3}) and $\beta = 0.39$ 
(\chem{YTiO_3}) are significantly below the mean field expectation of $0.5$ and 
lie within the range $0.3$-$0.4$ consistent with 3D critical fluctuations 
(e.g.~$0.346$ (3D~$XY$) or $0.369$ (3D Heisenberg))~\cite{pelissetto02}. This is 
reasonable in the context of the relatively isotropic nature of the exchange interactions 
in these compounds.

In the context of the dipole field calculations described in Section~\ref{sec:xto:dipole}
and the previous literature, the sites obtained for the two compounds considered 
here seem entirely plausible. For both compounds we find a site corresponding to the 
lower precession frequency in the \chem{A}--\chem{O} layer, as in \chem{LaMnO_3}, but the 
origin of the higher frequency component is almost certainly different in the two cases. 
In \chem{LaTiO_3} the higher frequency sites also appear to be in the rare earth--oxygen 
layer, and this fits with the equal amplitudes of the two components observed in the \musr{} 
signal. Sites near the \chem{Ti}--\chem{O} planes seem unlikely on the basis of the calculations.
In \chem{YTiO_3} the higher frequency component cannot be in the \chem{Y}--\chem{O} plane if the hyperfine coupling is negligible. A more plausible assignment corresponds to sites lying between two oxygen ions and relatively close to the \chem{Ti} ions, which would explain the high precession frequency, the relatively small amplitude (since the site would probably be less electrostatically favourable), and also the large initial phase offset, consistent with a stronger coupling to the 
antiferromagnetically coupled moments in the $ab$-plane. Lower frequency sites could also 
occur in the \chem{Ti}--\chem{O} layers. In both compounds a full site determination would 
require measurements on single crystals and in applied fields, as was done for 
the rare earth orthoferrites~\cite{holzschuh83,boekema84,lin86} and \chem{LaMnO_3}~\cite{cg01}. 

In magnetically ordered polycrystalline samples we would expect the relaxing component 
to account for around one third of the relaxing asymmetry, owing to the polycrystalline 
averaging of the effects of the magnetic fields parallel and perpendicular to the muon 
spin direction. 
The situation in these materials is not this straightforward. The fast initial relaxation 
$\sigma_{\rm r}$ is most likely to originate from large magnetic fields at muon stopping 
sites which are slightly magnetically inequivalent. The dipole field calculations suggest that
both compounds have plausible stopping sites close to the magnetic moment directions of nearby 
\chem{Ti^{3+}} ions, where a small range of muon stopping positions would give sufficiently 
different magnetic fields to lead to this fast relaxing component. 
Because of this rapid depolarization we are unable to distinguish the relaxation due to fields 
parallel to the muon spin direction and the amplitude $P_{\rm r}$ is likely to include the longitudinal relaxing component usually observed in polycrystalline magnets as well as the 
contribution from muons stopping at sites with very high local fields.

The results presented in this paper are in excellent agreement with previous reports of the 
magnetic properties of both \chem{LaTiO_3} and \chem{YTiO_3} obtained using neutron 
diffraction~\cite{PhysRevLett.85.3946,PhysRevLett.89.167202}. This confirmation is 
worthwhile given the history of sample dependent results and the difficulty of controlling 
the oxidation state precisely~\cite{PhysRevLett.97.157401,PhysRevB.71.184431}. Comparison 
between the precession frequencies observed in \chem{LaTiO_3} and dipole field calculations 
strongly favours moment alignment along the $a$-axis rather than the $c$-axis, an issue powder 
neutron diffraction has difficulty resolving~\cite{PhysRevB.68.060401}. 
Using a microscopic probe gives an independent means of testing the previous results from bulk 
probes; our results confirm that despite the complexities of the underlying orbital physics, 
both compounds behave magnetically as bulk, three-dimensional magnets. We are also able to test 
the ability of dipole field calculations to reproduce the magnetic field distributions within 
oxide materials. This is successful for these compounds, where the similarity of both the 
structure and the muon sites nevertheless yields different internal fields due to the 
significantly different magnetic structures. 

\section{Acknowledgements}
\label{sec:xto:acknowledgements}

Part of this work was carried out at the Swiss Muon Source, Paul Scherrer Institute, Villigen,
Switzerland. We thank Alex Amato for technical assistance. This research project has
been supported by the EPSRC and by the European Commission under the 6th Framework
Programme through the Key Action: Strengthening the European Research Area, Research
Infrastructures, Contract no R113-CT-2003-505925. 
TL acknowledges support from the Royal Commission for the Exhibition of 1851.

\end{document}